\documentclass[a4paper]{jpconf}
\usepackage{graphicx}
\usepackage{ragged2e}
\begin{document}
\title{Co-optimization of a piezoelectric energy harvesting system for broadband operation}

\author{S Zhao$^1$, U Radhakrishna$^2$, S Hanly$^3$, J Ma$^4$, J H Lang$^2$ and D Buss$^5$}

\address{$^1$ School of Microelectronics, Tianjin University, 
	Tianjin 300072, China}
\address{$^2$ Department of Electrical Engineering \& Computer Science, 
	Massachusetts Institute of Technology, Cambridge, MA 02139, USA}
\address{$^3$ Mide Technology, Medford, MA 02155, USA}
\address{$^4$ School of Computers, 
	Guangdong University of Technology, Guangzhou 510006, China}
\address{$^5$ Texas Instruments, Dallas, Texas 75243, USA}

\ead{shengzhao@tju.edu.cn}

\begin{abstract}
The goal of this research is to increase the bandwidth (BW) over which substantial energy can be harvested using a piezoelectric energy harvester (PEH). The key innovation is the use of bias-flip (BF) electronics at the output of a PEH having a large electromechanical coupling coefficient $\kappa_{e}^{2}$. For a PEH with large $\kappa_{e}^{2}$, the open-circuit resonance frequency $f_{oc}$ is substantially larger than the short-circuit resonance frequency $f_{sc}$. Over the intervening range, the reactive part of the conjugate matched load impedance is small, and can be approximated using BF electronics in which the BF voltage is sufficiently small and the BF losses are small. This results in a large BW over which substantial energy can be harvested. Experimental results using a commercially available PEH are presented to demonstrate this concept.  Design guidelines are provided for achieving PEHs having increased $\kappa_{e}^{2}$.
\end{abstract}

\section{Introduction}
Kinetic energy harvesters have become a viable source of electric power for low-power wireless sensor networks that might be connected to the Internet of Things. Of the many harvesters through which kinetic energy can be converted to electric energy, PEHs are often favored for their simple structure, their high power density, their high output voltage, and their ease of self starting. However, for the common low-loss resonant PEHs, the harvested power drops off dramatically when the ambient vibration frequency deviates from the resonance frequency of the PEH; only at the resonant frequency can the PEH output maximum power. This is problematic when PEH manufacturing tolerances and/or uncertainties in the vibration spectrum result in a frequency mismatch. In such cases it is desirable to increase the energy harvesting BW.

Several nonlinear interface circuits such as SSHI, BF, SECE and SSDCI have been proposed to increase the harvested power [1-5]. These techniques were demonstrated to be efficient at the resonance frequency of the PEH. At frequencies away from resonance, it was subsequently shown that BF employing universal phase can effectively optimize output power [6]. The BF technique shares features that are common with the independently developed P-SSHI-$\phi$ technique [7]. Even so, for a PEH with a small coupling coefficient ($\kappa_{e}^{2}$), the BF losses limit output power away from resonance because the BF voltage for frequencies away from resonance becomes large. Therefore, to compensate for the limited output power, this paper studies the resonance splitting behaviour of PEHs with large $\kappa_{e}^{2}$, which was initially explored in [8] with only resistive load. Resonance splitting with large $\kappa_{e}^{2}$ gives two well-separated output power peaks between which (and slightly beyond) BF electronics can harvest near peak power, thereby extending the harvesting BW. With a combination of BF employing universal phase, and a PEH having large $\kappa_{e}^{2}$, the energy harvesting BW can be further increased. Experimental results with a commercially-available PEH having $\kappa_{e}^{2}$=0.069 are presented to demonstrate this concept. With the help of BF, a peak power of 112 $\mu$W, and a 3-dB BW of 50 Hz over which substantial energy is harvested, are achieved. Additionally, this paper provides design guidelines for achieving PEHs having larger $\kappa_{e}^{2}$, and a new PEH is suggested to improve the BW performance of the existing commercial PEH.
\begin{figure}[t]
	\makeatletter\def\@captype{figure}\makeatother
	\begin{minipage}{.35\textwidth}
		\begin{center}	
			\includegraphics[width=11pc]{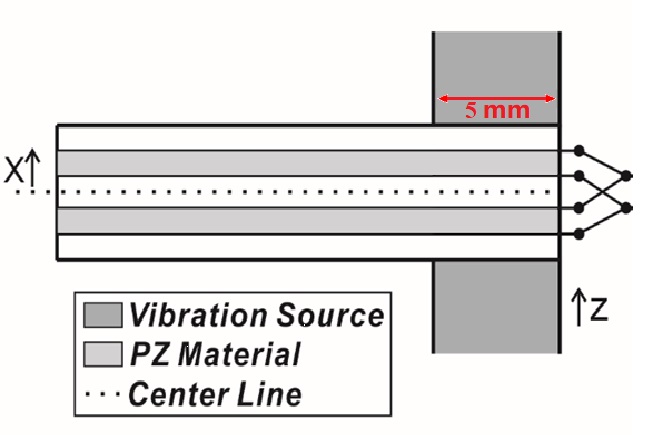}
			\caption{\label{label}Schematic of a cantilever PEH.}
		\end{center}	
	\end{minipage}
	\hspace*{.02\textwidth}
	\makeatletter\def\@captype{table}\makeatother
	\begin{minipage}{.6\textwidth}
		\caption{\label{jfonts}Four4 frequencies that are important to understanding a cantilever PEH.}
		\begin{tabular}{@{}l*{15}{l}}
			\br
			Value (Hz)&Definition\\
			\mr
			$f_{sc}$=673.00&short-circuit resonance frequency\\
			$f_{ZR1}$=673.43&lower zero Thevenin reactance frequency\\
			$f_{ZR2}$=695.72&upper zero Thevenin reactance frequency\\
			$f_{oc}$=696.00&open circuit resonance frequency\\
			\br
		\end{tabular}
	\end{minipage}
\setlength{\belowcaptionskip}{-1cm}
\end{figure}

\section{Characterization of a Commercial PEH}
The energy harvesting approach proposed here is studied experimentally with a commercial PEH. Mide PPA2014 PEH was selected because of its large $\kappa_{e}^{2}$. Figure 1 shows a cross section of this cantilever device. Figure 2a shows the equivalent circuit for the PEH that defines the compact model (CM) parameters. Extracted CM parameters and output voltage measurements made using 1-g acceleration and a 5 mm clamping overlap are shown in Figure 2b. Four frequencies are important to understand power output from a cantilevered PEH. They are explained in Table 1. The matched-load (ML) resistance $R_{ML}$ gives maximum output power at the lower zero-Thevenin-reactance frequency $f_{ZR1}$.

At the two frequencies $f_{ZR1}$ and $f_{ZR2}$, the reactive part of the impedance looking into the output terminals of the circuit model in Figure 2a is zero. At these frequencies the source force (voltage $V_{F}$) and velocity (current $I_{S}$) are in phase, and maximum power is delivered to a resistive load that is “matched” to the internal resistance of the device.

\section{Bias-Flip Circuit}
Figure 2c shows the DC rectification and storage (DCRS) circuit used here. The BF inductor, shown in the shaded area, is used to implement the tunable inductive or capacitive reactive impedance that is required to deliver maximum output power to the storage cell at a given vibration frequency. Operation of the BF or “switched” inductor has been described in detail elsewhere [1-3], and used in a variety of ways to increase output power from a PEH [4, 5]. Here, the bias, that is $V_{Out}$ across $C_{P}$, is flipped so as to produce a waveform having a phase $\phi(V_{Opt})$ in Figure 2d that would result if a matched reactive load were used in place of the BF inductor.
	
The BF timing is determined as follows. Begin with the AC Matched Load (ACML) circuit shown in Figure 2d. At each vibration frequency, the conjugate-matched impedance that delivers maximum real power to the matched resistive load is determined. With this impedance the phase $\phi(f)$ of the voltage $V_{Out}$ relative to the source voltage $V_{F}$ is determined to be	
\begin{equation}
\phi(f)=tan^{-1}\left [({1}/{(2\pi f C_{m})}-{2\pi f L_{m}})/{R_{m}} \right ]
\end{equation}

The bias-flipped voltage $V_{Out}$ is then flipped at the phase $\phi(f)$ relative to $V_{F}$. Despite the BF timing, to achieve maximum output power, in ACML case the matched resistive load $R_{L}$ is required, while for DCRS case an optimum rectification voltage $V_{Rect}$ is required.

It is well-known that the maximum output power from a PEH is given by 
\begin{equation}
P_{Opt}={m^2a^2}/{(8 \eta)}={V_{F}^2}/{(8R_{m})}
\end{equation}

For the ACML circuit, this maximum output power is achieved at all frequencies. The black dashed curve of Figure 4a shows that, if the conjugate matched reactive load is replaced by a BF circuit having 100\% BF efficiency (no inductor loss), and the resistive load is replaced by an ideal rectifier having the proper $V_{Rect}$, the power delivered to the storage cell is within 5\% of $P_{Opt}$ over a wide frequency range. As will be seen in Figure 3, the $V_{Out}$ waveform in the BF/DCRS circuit looks very different from the sinusoidal voltage in the ACML circuit. In spite of this, ACML output power is a good predictor for BF/DCRS output power.

\begin{figure}[t]
	%	\makeatletter\def\@captype{figure}\makeatother
	\begin{center}
		\includegraphics[width=.8\textwidth]{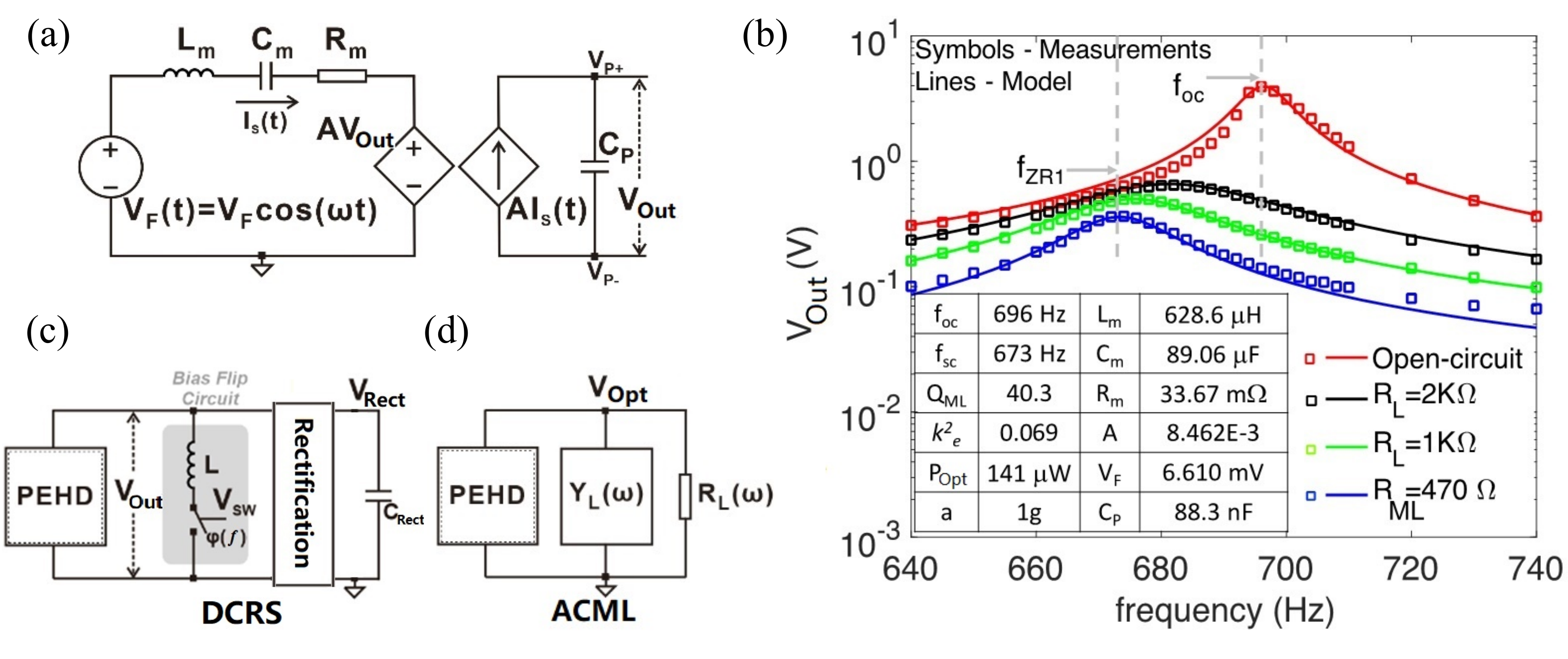}
		\caption{\label{label}(a) Equivalent circuit of a PEH, (b) CM parameters, simulated and measured voltage curves of the Mide PPA2014 PEH for various resistive loads with a 5 mm clamping overlap, (c) DCRS circuit and (d) ACML circuit.}
	\end{center}
\end{figure}

A limitation of the BF technique is that its efficiency is typically between 80\% and 90\% [6]; the losses occur in the inductor and switch. When a PEH with low $\kappa_{e}^{2}$ is used, the BF voltage for frequencies away from resonance becomes large, and BF losses limit output power [3, 7]. However, for PEHs having large $\kappa_{e}^{2}$, $f_{sc}$ and $f_{oc}$ become well separated, the BF voltage in this range becomes low, and BF losses become small, thereby increasing the BW over which substantial energy can be harvested. The low BF voltage in this range is illustrated by the simulations in Figure 3. This behavior can be understood by analogy to the ACML circuit. For each vibration frequency $f$, there is a matched load resistor that gives optimum output power, and this resistor defines a resonance frequency $f_{res}$. Between BF pulses, the voltage oscillates at $f_{res}$. For the region between $f_{sc}$ and $f_{oc}$, $f_{res}$ is close to the vibration frequency $f$. Because of this, a very small tweak is required by the BF circuit to achieve the correct phase of $V_{Out}$. Far above $f_{oc}$ and far below $f_{sc}$, there is a large difference between $f_{res}$ and $f$, and a large BF voltage is required. In summary, the combination of BF and large $\kappa_{e}^{2}$ can be used to significantly extend the range over which a PEH can produce (near) maximum power output without the need for high-voltage loading electronics.

\section{Experimental and Simulation Results}
To demonstrate widening of the frequency range over which (near) maximum power can be harvested, experiments and simulations are performed with the Mide PPA2014 PEH described in Section 2. A 1-g acceleration and a 5 mm clamping overlap are used. The results shown in Figure 4 are based on the DCRS circuit shown in Figure 2c in which the rectifier is a diode bridge. Output power is optimized by varying $V_{Rect}$ at each frequency. The blue curve shows that, even with 82\% BF efficiency, the output power is very near the theoretical maximum (green curve) between $f_{sc}$ and $f_{oc}$, given ideal diodes. Diode voltage drop degrades output power (red curve) but the resulting 3-dB BW of 50 Hz (7.4\% of $f_{sc}$) represents a substantial increase over a PEH with negligible coupling, where the 3-dB BW is $f_{sc}/Q_{ML}$=17Hz (2.5\% of $f_{sc}$).

\begin{figure}[t]
%	\makeatletter\def\@captype{figure}\makeatother
	\begin{center}
		\includegraphics[width=.95\textwidth]{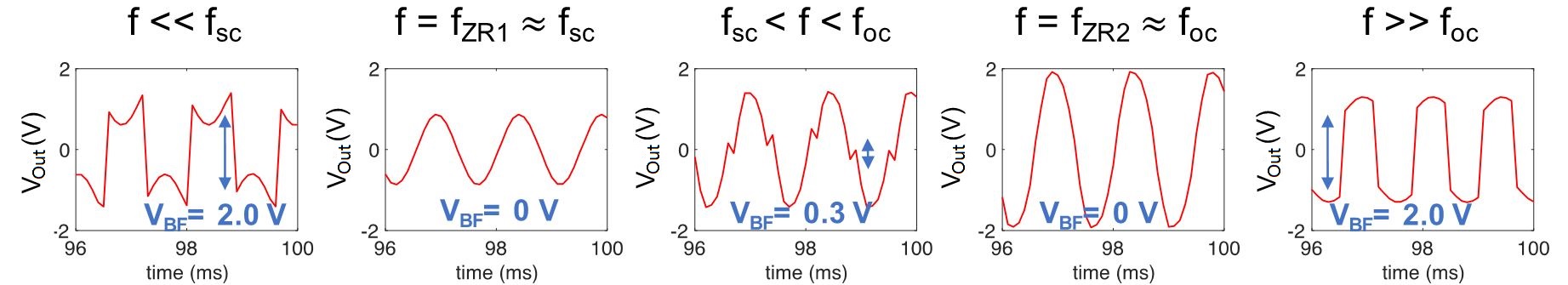}
		\caption{\label{label}Waveforms of $V_{Out}$ for selected vibration frequencies. For $f \ll f_{sc}$ and $f \gg f_{oc}$, the BF voltage $V_{BF}$ is large, and BF loss is significant. However, for $f_{sc} < f < f_{oc}$ $V_{BF}$ is small. At the two zero-Thevenin reactance frequencies $f_{ZR1} \approx f_{sc}$ and at $f_{ZR2} \approx f_{oc}$, $V_{BF}=0 V$.}
	\end{center}	
\end{figure}
\begin{figure}[b]
%	\makeatletter\def\@captype{figure}\makeatother		
	\begin{center}	
		\includegraphics[width=.85\textwidth]{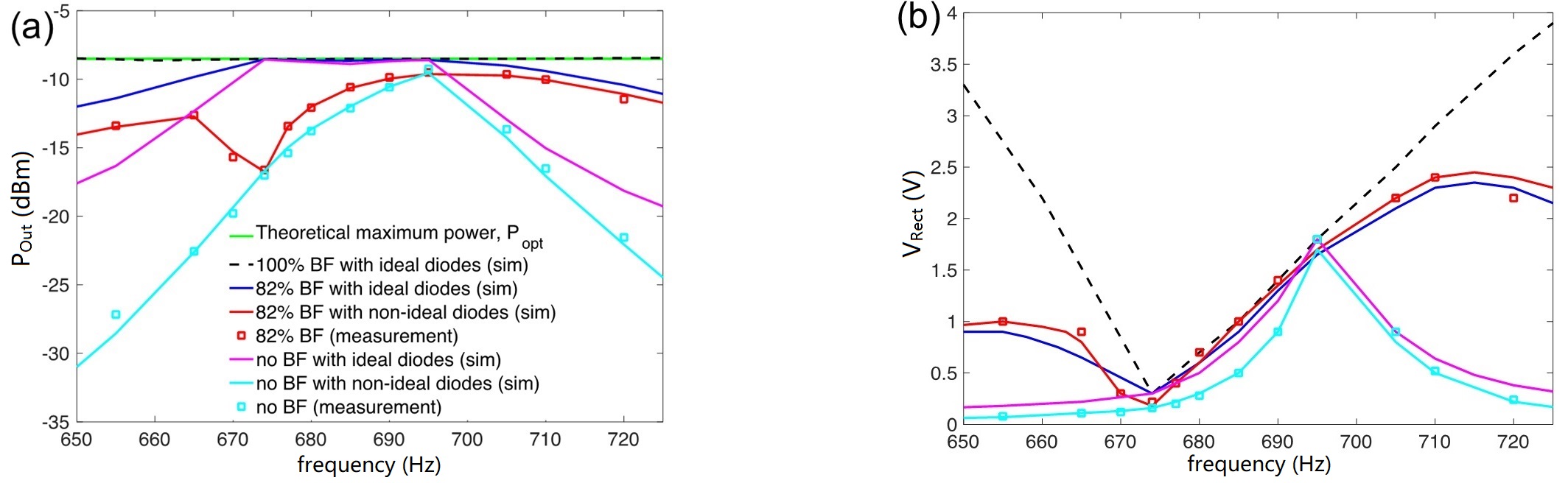}
		\caption{\label{label}(a) Maximum output power and (b) corresponding value of $V_{Rect}$ that gives maximum output power for a variety of load conditions.}	
	\end{center}
\end{figure}

Figure 5 shows the effect of varying $\kappa_{e}^{2}$. These simulations show output power for PEHs having the CM parameters shown in Figure 2b, but with the parameter A varied to provide different $\kappa_{e}^{2}$, where A represents the coupling between mechanical and electrical energy in a PEH. The simulations are made assuming a realistic 82\% BF efficiency. They further assume the use of a “smart” switching rectifier in the DCRS/BF circuit of Figure 2c to eliminate the diode-voltage loss around $f_{sc}$. When $\kappa_{e}^{2}$ is small, maximum output power occurs at $f_{ZR1}$$\approx$$f_{sc}$. As $\kappa_{e}^{2}$ increases, the 3-dB BW increases. Use of the “smart” switching rectifier increases the BW with the present device ($\kappa_{e}^{2}$=0.069) to 78 Hz (11.5\% of $f_{sc}$); if $\kappa_{e}^{2}$ can be increased by a factor of 2, the 3-dB BW increases to 104 Hz (15.4\% of $f_{sc}$).

\section{Design for Large $\kappa_{e}^{2}$}
The coupling coefficient $\kappa_{e}^{2}$ is given by [9]
\begin{equation}
\kappa_{e}^{2}={A^2C_{m}}/{C_{P}}={A^2}/{(kC_{P})}={A^2}/{((k_{PE}+k_{non-PE})C_{P})}={(f_{oc}^2-f_{sc}^2)}/{f_{sc}^2}
\end{equation}where $C_{m}$ represents the beam stiffness, $C_{P}$ represents the plate capacitance of the piezoelectric material, $k$ is the total spring constant of a PEH, $k_{non-PE}$ and $k_{PE}$ are spring constants of non-PE and PE component respectively. For the bimorph cantilever of Figure 1, A is given by [10]
\begin{equation}
A={3WY_{PE}b_{PE}d_{31}}/{L}
\end{equation}
where W and L are width and length of a PEH, $b_{PE}$ is the distance from the center-line of the cantilever to the center-line of the piezoelectric (PE) layer, $Y_{PE}$ is the Young’s modulus of PE material, and $d_{31}$ is the piezoelectric coefficient.

\begin{figure}[t]
%	\makeatletter\def\@captype{figure}\makeatother
	\begin{minipage}{.4\textwidth}
		\begin{center}	
			\includegraphics[width=.95\textwidth]{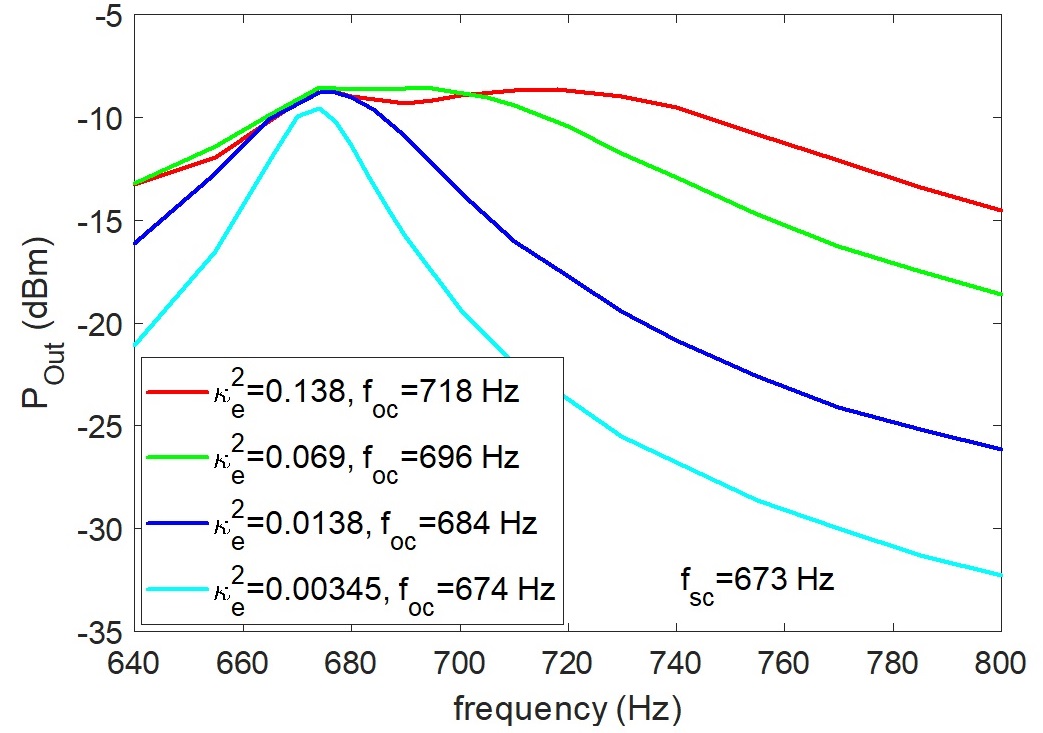}
			\caption{\label{label}Output power for PEHs having different $\kappa_{e}^{2}$.}
		\end{center}  	
	\end{minipage}
	\hspace*{.03\textwidth}
%	\makeatletter\def\@captype{figure}\makeatother
	\begin{minipage}{.55\textwidth}
		\begin{center}	
			\includegraphics[width=.84\textwidth]{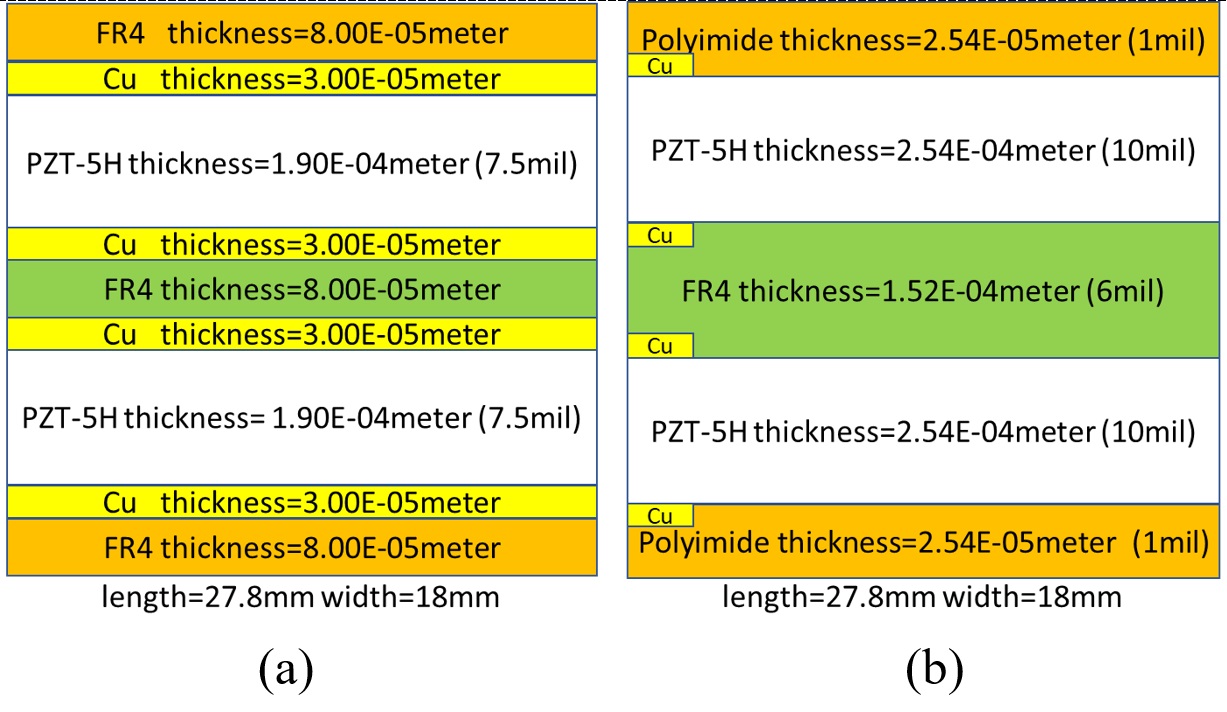}
			\caption{\label{label}(a) Cross section of baseline device and (b) proposed redesign.}
		\end{center}  
	\end{minipage}
\end{figure}

A and $C_{P}$ depend only on the PE material. It is apparent that $k_{non-PE}$ should be made as small as possible. Thus, a composite cantilever designed for large $\kappa_{e}^{2}$ should be designed such that the electrodes and non-PE outer protection layers have low Young’s modulus, and are as thin as possible. But the Young’s modulus of non-PE material shouldn’t be too small such that shearing occurs and two PE layers vibrate independently. Additionally, increasing the coupling term A also leads to a large $\kappa_{e}^{2}$. If the PE material is predetermined in the design, increasing $b_{PE}$ is a good choice to increase A. While changing of W or L is not suggested since not only A but also $C_{P}$ and $k$ are related to W and L. To increase $b_{PE}$, thick non-PE material should be inserted in between two PE layers, and the device center-line has minimal impact on spring constant.

Figure 6a shows the cross section of the commercial PPA2014. The non-PE material in this device contributes 41.6\% of the total spring constant, and the $\kappa_{e}^{2}$ is calculated to be 0.053 (0.069 measured) with 5 mm clamping overlap. A redesigned device is shown in Figure 6b. The redesigned device retains the original L and W. To reduce $k_{non-PE}$, the Cu electrodes are removed, and the FR4 outer layer is replaced by polyimide which has a lower Young’s modulus. Thicker PE capacitors are adopted to increase $k_{PE}$. In addition, a thicker center FR4 layer is used to increase $b_{PE}$ so that A is increased. In the redesigned device, with a 5 mm clamping overlap, the non-PE material contributes 5.6\% of the spring constant, and $\kappa_{e}^{2}$ is predicted to be increased by 2x to $\kappa_{e}^{2}$=0.121.

\section{Conclusions}
One factor that has prevented the widespread commercial use of PEHs is the narrow range of vibration frequencies over which (near) optimum power can be harvested. This paper shows that, with a combination of bias-flip electronics, smart rectification, and a PEH optimized for large $\kappa_{e}^{2}$, one can achieve a fractional 3-dB BW in excess of 15\%.

\end{document}